\documentclass[11pt,twoside,onecolumn]{article}
\usepackage[]{latexsym,amsmath,amssymb}
\usepackage[]{epsfig}
\newcommand{\be}{\begin{eqnarray}}
\newcommand{\ee}{\end{eqnarray}}

\renewcommand{\H}{\hat H}
\newcommand{\p}{\hat p}
\renewcommand{\P}{\hat{\cal P}_\varphi}

\title{\bf A quantum-like description of the planetary systems}
\author{Fabio Scardigli\thanks{E-mail: fabio@yukawa.kyoto-u.ac.jp}\\
{\em Yukawa Institute for Theoretical Physics,}\\
{\em Kyoto University, Kyoto 606-8502, Japan}}
\date{}
\begin{document}
\maketitle
\begin{abstract}
The Titius-Bode law for planetary distances is reviewed. A model
describing the basic features of this rule in the "quantum-like"
language of a wave equation is proposed. Some considerations about the 't Hooft idea
on the quantum behaviour of deterministic systems with dissipation are discussed.\\
%
%
%\par
%\null
%\par
$\quad$\\
\textit{Keywords: Foundations of Quantum Mechanics, Planetary
Systems. PACS: 03.65.Ta, 96.35.-j}
\end{abstract}
\raggedbottom
\setcounter{page}{1}
%
%
%
%$\quad$\\
%\textsf{And I cherish more than anything else the Analogies,\\
%my most trustworthy masters. They know all the secrets of Nature,\\
%and they ought least to be neglected in Geometry.
%\\
%Johannes Kepler}
%
%
%
\section{Introduction and motivation}
\setcounter{equation}{0}
In the last years many papers, in particular those of 't Hooft
\cite{hooft1}, but also of several other authors \cite{kleinert, elze, blasone},
have described a
possible way out from some of the unpleasant (or, to cite
Feynman, "peculiar") features of Quantum Mechanics. \\
The idea proposed in \cite{hooft1} is, in a nutshell, that a classical
deterministic theory supplemented with a
dissipation mechanism (information loss) should produce at larger
scales a quantum mechanical behaviour. The
emergent model, after that the dissipation mechanism has been
implemented, can be described with the usual tools of Quantum
Mechanics, namely,  states evolving in Hilbert space, unitary
matrix, wave equation, etc.\\
Originally motivated (also) by the challenge of Quantum Gravity,
the initial accent on the Planck scale collocation of the deterministic
underlying theory has been more recently weakened in favour  of
the description of several classes of deterministic classical
systems, where quantum behaviour emerges from classical dynamics
constrained by dissipation mechanism(s) (see \cite{kleinert, elze, blasone}).
So far, no explicit scale has been implemented in such models.\\
On the other hand, it is quite clear that dissipative/chaotic processes
(of energy, angular momentum, etc.) have acted, and act, also on scales very far
from the Planck scale. For example, in the proto-planetary nebulae there
are many dissipative and chaotic processes due to friction, viscosity, gravitational
resonances, etc. Therefore, it is worth asking if these dissipative/chaotic processes
could produce, on different scales, results
analog to those predicted by the 't Hooft theory, namely some sort of quantum-like behaviour.
Surprisingly enough, Nature seems to play just in this way. The Titius-Bode law of planetary
distances seems to bring evidence along this direction.
In analogy with the 't Hooft idea, the chaotic dissipation occurred during the formation
period of a planetary system could have produced {\em "the
apparent quantization of the orbits which resembles the structures
seen in the real world"} \cite{thooft2}.
In the present paper, an "effective" quantum-like kinematics
(in the form of a deformed Schr\"odinger wave equation) is developed and applied to the
description of planetary systems. In doing this, we obtain
also a quite precise prediction of the internal radius of the rings surrounding
Saturn, Jupiter and Uranus.
In the end, some qualitative considerations about permitted orbits and dissipative chaotic
mechanism(s) are discussed.
\section{Titius-Bode law}
\setcounter{equation}{0}
The Titius-Bode law is the most famous among the relationships about the spacing of planetary systems.
It is an empirical rule which gives, under the
hypothesis of circular orbits, the distances of the planets from
the Sun as a function of a single parameter, i.e. an integer $n$. There
are several versions of the law. The eldest form is perhaps the
following
\be
r(n) = 0.4 + 0.3 \cdot 2^n
\ee
where $r(n)$ is given
in Astronomical Unit (1 A.U. $\simeq 150 \cdot 10^6$ km). For
$n=-\,\,\infty, 0, 1, 2,...$ this rule gives the distances,
respectively, of Mercury, Venus, Earth, Mars, etc., including the
asteroids belt (actually, Cerere was discovered following the
indications of this law) and Uranus, which at the moment of the
first formulation of the law (1766-1772) had not yet been
discovered (see the book of Nieto \cite{nieto} for history, explicit
formulations, theories and extensive comments).\\
In this original formulation, the law was not able
to account for the distance of Neptune and Pluto. The more recent
versions of the rule, elaborated during the XX century by Blagg (1913) and Richardson
(1943), are able to describe not only the planetary
distances within the solar system, including planets like Neptune
and Pluto, but also can be successfully applied to the systems of
satellites orbiting Jupiter, Saturn and Uranus. The agreement
between the predicted and the observed distances of the various
satellites from the central body is really good, of the
order of a few
percents, as can be checked in the tables of Nieto. Of course, criticisms have been risen even against
an effective physical meaning of the law (see, for example, \cite{dubrulle}). However,
in \cite{lynch}, it is also convincingly argued  that the agreement with the observations
cannot be safely considered as a mere statistical chance.
This modern version of the Titius-Bode-Richardson rule seems then to have
a quite firm status, certainly not yet comparable to that of the
Balmer formula for the atomic spectra, but at least similar.\\
The main feature of these versions of the
Titius-Bode law is that the rule can be expressed, if we neglect
second order corrections, by an exponential relation as
\be
r=a\,e^{2\lambda n},
\label{tbl}
\ee
where the factor $2$ is
introduced for convenience reasons and $n=1,\, 2,\, 3,\, \dots$.\\
For the Solar System we have
\be
2 \lambda &=& 0.53707, \quad \quad \quad  e^{2\lambda} \simeq 1.7110, \nonumber \\
a &=& 0.21363 \,\,\, {\rm A.U.} \nonumber
\ee
The amazing thing found by Blagg was that
the geometric progression ratio $e^{2\lambda}$ is roughly the same
for the Solar System and for the satellite
systems of Jupiter ($e^{2\lambda} \simeq 1.7277$), Saturn ($e^{2\lambda} \simeq 1.5967$),
and Uranus ($e^{2\lambda} \simeq 1.4662$). The parameter $\lambda$ is dimensionless,
its value is taken from the observed data, for any of the systems considered
(planetary or satellite systems).
The parameter $a$ has the dimension of a length and it is linked to the radius of the first
orbit of the system considered (in fact $r(1)=ae^{2\lambda}$).\\
Many theories have been developed during the last 240 years
in order to explain, or describe, the Titius-Bode law.\\
There have been dynamical models connected with the theory of the
origin of the solar system, electromagnetic theories,
gravitational theories, nebular theories. They all can be found
in literature (see, for example, \cite{various}) and they have
been excellently reviewed in the book \cite{nieto}.
We remind that also the
idea of using a Schr\"odinger-type equation in order to give
account of the law (\ref{tbl}) is not new: during the years many authors
have suggested various approaches in this direction (see, for example, an incomplete list of
papers in \cite{sch})\footnote{However, in almost any of the papers of Ref.
\cite{sch}, the Newtonian potential $1/r$ is usually
assumed together with a Bohr-like quantization
condition, and these two things together yield (as it is well known) a law for the
orbital radii as $r\sim n^2$ (not as $r\sim e^{2\lambda n}$). This forces the authors
either to use non consecutive integers to accomodate the planet distances, or to consider the solar system divided
in two subfamilies (i.e. terrestrial planets and gigantic planets).}\,.
Also the present paper aims to develop a quantum-like model of the T-B rule,
but this will be done through a deformation of the usual Schr\"odinger equation.
\section{Bohr model of the hydrogen atom}
\setcounter{equation}{0}
In this section we remind briefly the Bohr model for the hydrogen atom, just to set the notation.
The electron orbits are supposed circular (this will be held for
planetary orbits also). The two main equations are the equation
for the force (i.e. the equation of motion)
\be
\textrm{m} \frac{v^2}{r} = \frac{{\rm e}^2}{r^2}
\ee
where "$\textrm{m}$" and "$\rm{e}$" are the mass and charge of the electron; and the
quantization condition on the ($z$-component) of the angular
momentum
\be
\textrm{m}vr = n \hbar \quad \quad \quad n=1,2,3,...\,.
\label{bohr}
\ee
In the Bohr model, all the orbits belong to the same plane, and
this is also taken for true in the planetary models. From the two
equations above, one easily derives
\be
\left\{ \begin{array}{l}
r=\dfrac{\hbar^2}{\textrm{m}\,{\rm e}^2}\,n^2
\\
\\
v=\dfrac{\hbar \, n}{\textrm{m}\,r}
\end{array}
\right.
\ee
The first equation is the law of electron distance from the nucleus in the Bohr model.
With this law, from the classical expression for the total energy we get the energy
spectrum of the bound orbits
\be
E= \frac{1}{2} \textrm{m} v^2 - \frac{{\rm e}^2}{r} = -\frac{\rm{e}^2}{2r} =
-\frac{\textrm{m}{\rm e}^4}{2\hbar^2}\,
\frac{1}{n^2}
\ee
Now we shall try to apply analog ideas to the solar and planetary systems.
\section{Model {\em a la} Bohr for a planetary system}
\setcounter{equation}{0}
\label{Sec4}
In this section we introduce a model for the "quantization" of a
planetary system. The model acquires its discrete, or "quantum", properties
from a modification of the Bohr quantization rule for the angular
momentum. The equations here proposed, for a generic planet of mass $\textrm{m}$, orbiting
a central body of mass $M$, are \be \left\{
\begin{array}{l} \textrm{m} \dfrac{v^2}{r} = \dfrac{GM\textrm{m}}{r^2}
\\
\\
\dfrac{J}{\textrm{m}}=vr = s \,e^{\lambda n}
\end{array}
\right.
\label{tbam}
\ee
where $n=1,2,3,...$, $s$ is a constant, and $J$ is the angular momentum of the planet.
Some comments are immediately required:
\begin{itemize}
\item Because of the principle of equivalence, the masses "m"
on the LHS and on the RHS of the first eq. (\ref{tbam}) cancel out each
other.
\item The constant $s$ in the RHS of the second eq.
(\ref{tbam}) has the dimensions of an action per unit mass. It
plays the role of $\hbar$ and it must be understood as an action
typical of the planetary system under consideration. It is not
possible to use $\hbar$ itself, because this would fix the wrong
initial radius in the Titius-Bode law, that is the constant
$a$ in $r=ae^{2\lambda n}$.
\item The constant $\lambda$ in (\ref{tbam}) is
the phenomenological parameter obtained from the observation
($2\lambda = 0.53707$ for the Sun, $2\lambda = 0.54677$ for Jupiter,
$2\lambda = 0.46794$ for Saturn,
$2\lambda = 0.38271$ for Uranus).
\item In the second of the eqs. (\ref{tbam}), we quantize the
angular momentum {\em per unit mass} (i.e. the velocity field and not the momentum field).
This a consequence of the principle of equivalence. If we did not do so,
we would obtain a law for $r(n)$ where the scale of distance
changes from a planet to another, as the planetary masses change from a planet to another.
\end{itemize}
From (\ref{tbam}) one immediately gets
\be
v= \frac{s\,e^{\lambda n}}{r} \quad \quad \Rightarrow \quad \quad
r(n)= \frac{s^2}{GM}\,e^{2\lambda n}
\label{rtbam}
\ee
which is the Titius-Bode law if we identify $a=s^2/(GM)$.\\
We can also compute the energy spectrum for the i-th planet from (\ref{rtbam})
\be
E_{(i)}=
\frac{1}{2}\textrm{m}_i v^2 - \frac{GM\textrm{m}_i}{r}
= -\frac{GM\textrm{m}_i}{2r} =
-\left(\frac{GM}{s}\right)^2\frac{\textrm{m}_i}{2e^{2\lambda n}}\,,
\ee
where $n=1,2,3,\dots$. As we see, the energy of the i-th planet is
not properly quantized by itself. This is because the mass $\textrm{m}_i$
changes in general with the planet, and this implies different
sets of energy levels for different planets. Instead, the energy
per unit mass
\be
{\cal E}
:=\frac{E_{(i)}}{\textrm{m}_i}=-\left(\frac{GM}{s}\right)^2\frac{1}{2e^{2\lambda n}}
=-\frac{1}{2}\,\frac{GM}{a}\,\frac{1}{e^{2\lambda n}}
\label{etbam}
\ee
is exactly quantized, i.e. it is a quantity
which depends on the orbit $n$ only (apart from the general constants $G$,
$M$, $s$), and not on the mass of the planet belonging to that orbit.
Therefore, the energy-per-unit-mass levels are valid for
the whole set of orbits of a specific planetary system.\\
\\
Also here some comments are needed, in order to complete the
explanation given before.
\begin{itemize}
\item The constant $s$ can be
computed in terms of the mass $M$ of the central body and of the
parameter $a$ (remind that the radius of the first orbit is
$r_1 = ae^{2\lambda}$)
\be
s=\sqrt{GMa}
\ee
\item The constant $s$ is {\em not} the same for all the
planetary systems (Sun, Jupiter, Saturn, Uranus). In fact, if it
were so, this would imply that the parameter $a=s^2/(GM)$ should
be in inverse proportion to the mass $M$ of the central body,
which is not true. Therefore the constant $s$ is not universal,
like $\hbar$, but it depends on the planetary system under
consideration (it changes from planetary system to planetary system).
\item In the Titius-Bode law the mass of the single planet (or satellite) plays no
role, essentially because of the principle of equivalence. A naive
quantization \emph{a la} Bohr, with a quantum condition $\textrm{m}vr=se^{\lambda n}$,
would yield a system of orbits with radii
\be
r(n)=\frac{s^2}{GM \textrm{m}_i^2}\,e^{2\lambda n}
\ee
(and corresponding energy levels). In such a system, every specific i-th planet, with
mass $\textrm{m}_i$,
would have its own system of permitted orbits, and the coefficient $a=s^2/(GM \textrm{m}_i^2)$ of
the T-B law would depend on the mass of the specific planet under consideration. All this is
contrary to the observed evidence. In a planetary system the coefficient $a$ of the T-B law
is the same for all the planets belonging to such system. (Of course, $a$ changes from
planetary system to planetary system). Agreement with the observed evidence can be obtained if
we "quantize" the angular momentum \emph{per unit mass}, $J/\textrm{m}=vr=se^{\lambda n}$.
\end{itemize}
The same considerations hold for the energy levels. The set of energy levels of the i-th
planet is given by
\be
E_{(i)}=-\frac{GM \textrm{m}_i}{2 a e^{2\lambda n}}.
\ee
This implies that every planet has its own set of energy levels.
Therefore, in order to have a
quantity linked to the orbit only, and not to the mass of the planet belonging to that
orbit, we should consider the energy per unit mass, $\mathcal{E}=E_i/\textrm{m}_i$. \\
The difference between a planetary system and a muonic/electronic atom is in this respect
striking. In a muonic atom the levels belonging to the muon can be occupied by the muon only;
the electron has different levels. In a planetary system the orbit is singled out by the
orbital velocity only. Jupiter can be put in the orbit of the Earth, if to Jupiter is given
the same speed of the Earth. This is a simple consequence of the principle of equivalence.
\section{Wave equation for a planetary system}
\setcounter{equation}{0}
The Bohr-like model introduced in section \ref{Sec4} allows us to look for
the corresponding Schr\"{o}dinger-like equation.
On \emph{postulating} the absence of interference phenomena
among the probability amplitudes of the planets along their orbits, and of course the absence of quantum jumps
(i.e. transitions) among the orbits,
we shall now construct a wave equation, and we shall see what it predicts for
the energy spectrum and for the exponential positions of the
planets, when the usual probabilistic interpretation of the wave
function $\psi(\vec{x},t)$ is adopted.\\
We suppose, hence, to
associate to a planetary system a scalar field $\psi(\vec{x},t)$,
the so called wave function. We shall see that the wave function
does not give us information on the behaviour of the single
planet, but rather on the global structure (orbits, energy levels, rings, etc.)
of the whole planetary system (about this, see the papers of Reinisch and Nottale in \cite{sch}).\\
\\
On comparing the Bohr quantization condition $\textrm{m}vr=n\hbar$ with the
condition given in (\ref{tbam}), $vr=se^{\lambda n}$, we see that the most straightforward
correspondence is
\be
s \longleftrightarrow \frac{\hbar}{\textrm{m}} \quad
\quad {\rm or} \quad \quad \hbar \longleftrightarrow s\textrm{m}.
\ee
This correspondence allows us to write down immediately a wave equation
for the stationary states. In fact, from the Schr\"{o}dinger
eigenvalue equation
we can write
\be
\left[-\frac{s^2\textrm{m}}{2}\nabla^2+U(r)\right]\psi=E\psi \, ,
\ee
and defining
\be
\cal E &:=& \frac{E}{\textrm{m}} = {\rm energy\,\, per\,\, unit\,\, mass}  \nonumber\\
V(r)&:=&\frac{U}{\textrm{m}}= {\rm potential\,\, energy\,\, per\,\,
unit\,\, mass}
\ee
we arrive at
\be
\left[-\frac{s^2}{2}\nabla^2+V(r)\right]\psi=\cal E\psi \, ,
\label{esch}
\ee
which will be the starting point for the construction of our wave equation
for planetary systems.\\
We note again that the quantity
correctly quantized is $\cal E$, the energy per unit mass, and not
the energy itself. As already said for the problem {\em a la
Bohr}, this is a direct consequence of the principle of equivalence.
\section{Wave equation for the Titius-Bode problem}
\setcounter{equation}{0}
\label{sec9}
In this section we look for a wave equation corresponding to the
model (\ref{tbam}). We also take into account the observed fact that
all the planetary orbits lie (more or less) in the same plane by considering a
2-dimensional Schr\"{o}dinger-like equation, i.e. we write the
previous wave equation in a plane
(on this, see also M. de Oliveira Neto \emph{et al}. in Ref. \cite{sch}).
This choice will suggest to us how to modify the standard
Schr\"{o}dinger equation in order to accommodate for the particular quantization condition (\ref{tbam})
on the angular momentum, namely $vr=se^{\lambda n}$.\\
The equation (\ref{esch}) can be written in
operatorial form
\be
\H_{\textsc{m}}\psi={\cal E}\psi
\ee
where $\H_{\textsc{m}}$ is the {\em Hamiltonian per unit mass}
\be
\H_{\textsc{m}}=\frac{\vec{p}\,^2}{2}+V(r) \label{H}
\ee
with the association
$\vec{p}  \leftrightarrow -is\,\vec{\nabla}$.\\
The Hamiltonian $H_{\textsc{m}}$, written in two dimensions and in planar polar coordinates, reads
\be
\H_{\textsc{m}}=\frac{1}{2}\left(\p_r^2 + \frac{\p_\phi^2}{r^2} \right) + V(r).
\label{Hs}
\ee
With the usual associations of ordinary quantum mechanics, in planar polar coordinates,
\be
\p_r^2 &\longrightarrow& -s^2\frac{1}{r}\frac{\partial}{\partial r}
\left(r \frac{\partial}{\partial r} \right) \nonumber \\
\p_\phi &\longrightarrow& -is\frac{\partial}{\partial \phi} \quad
\Longrightarrow \quad \p_\phi^2 \longrightarrow
-s^2\frac{\partial^2}{\partial \phi^2}
\ee
the Hamiltonian (\ref{Hs}) reads
\be
\H_{\textsc{m}} = -\frac{s^2}{2r^2}\left[r\frac{\partial}{\partial
r} \left(r \frac{\partial}{\partial r} \right) +
\frac{\partial^2}{\partial \phi^2} \right] + V(r).
\ee
The operator which accounts for the whole angular momentum is $\p_\phi$ (in fact, the
motion takes place in the plane $(x,y)$, that is, the only non zero
component of the angular momentum is $J_z$).
On the Hilbert space ${\cal L}^2([0, 2\pi])$, the operator
$\p_\phi$ is hermitian (self-adjoint) with eigenfunctions $e^{im\phi}/\sqrt{2\pi}$ and eigenvalues
$ms$, $m\in\mathbb{Z}$ ($\p_\phi\,e^{im\phi} = ms\,\,e^{im\phi}$).
This corresponds to the quantum condition ({\em a la} Bohr)
$vr=ms$ (see (\ref{bohr})). Instead, in order to accomodate as eigenvalues
the numbers $se^{\lambda m}$, and, of course, to do not spoil or
loose the eigenfunctions $e^{im\phi}$ and their Hilbert space
${\cal L}^2([0, 2\pi])$, we define the
operator
\be
\P \,e^{im\phi} := i\,e^{m\lambda}\,e^{im\phi}
\label{defP}
\ee
for every $e^{im\phi} \in {\cal L}^2([0, 2\pi])$,
$m\in\mathbb{Z}$, where $\lambda$ is the phenomenological parameter
given in (\ref{tbl}) ($2\lambda \simeq 0.53707$ for the Solar System).\\
Note that such a definition is directly dictated by the model (\ref{tbam}),
if one wants to reproduce the correct eigenvalues of the angular momentum-per-unit-mass
(as we learnt them from (\ref{tbam})).
The operator $\P$ can be defined to be linear (see Appendix), and being defined
on an orthonormal basis of ${\cal L}^2([0, 2\pi])$, is therefore
well defined on all ${\cal L}^2([0, 2\pi])$. In the Appendix we
show that the operator $\P$ verifies
\be
\P=ie^{-i\lambda\partial_\phi}.
\ee
Thus, we change the usual association $\p_\phi \rightarrow -is\partial_\phi$
into the new one $\p_\phi \rightarrow -is\P$.
With this \textit{new} definition of $\p_\phi$ we have
\be
\p_\phi
e^{im\phi} = -is\P e^{im\phi} = se^{-i\lambda\partial_\phi}e^{im\phi} = s e^{\lambda
m}e^{im\phi}\,.
\ee
The operator $\p_\phi$ so defined can be proved to be
self-adjoint (see Appendix).\\
Evidently, being the operator $\P$ constructed by $\lambda$,
it will acquire its full physical meaning in a framework where the constant $\lambda$ is derived
from a chaotic/dissipative mechanism at work in the proto-planetary nebula. In this direction,
contacts are expected, for example, with the formulations proposed
by Laskar, or von Weiz\"acker \cite{various}, or Willerding \cite{will}.\\
The new Hamiltonian operator per unit mass reads
\be
\H_{\textsc{m}} =
-\frac{s^2}{2r^2}\left[r\frac{\partial}{\partial r} \left(r
\frac{\partial}{\partial r} \right) + \P^2 \right] + V(r).
\ee
The wave function in these {\em planar polar coordinates} can be
written as
\be
\psi(r, \phi)=R(r)\Phi(\phi)
\ee
where the separation of variables is supposed, in order to proceed towards
a solution of the Schr\"{o}dinger-like equation. The normalization
condition is
\be
\int_{\mathbb{R}^2}|\psi(r,\phi)|^2 d^2 x
=\int_0 ^\infty R^2(r)r dr \cdot \int_0^{2\pi} |\Phi(\phi)|^2d\phi=1.
\label{norm}
\ee
The time-independent Schr\"{o}dinger-like equation is $\H_{\textsc{m}}\psi={\cal E}\psi$.
Since it is a 2-dimensional equation and it involves a deformed angular
momentum operator, we proceed now to work out the solutions explicitly.
With the position $\psi(r, \phi)=R(r)\Phi(\phi)$ the equation becomes
\be
\frac{r}{R}\frac{\partial}{\partial r}\left(r \frac{\partial
R}{\partial r} \right) + \frac{1}{\Phi}\P^2 \Phi +
\frac{2r^2}{s^2} ({\cal E}-V(r)) = 0 \, , \nonumber
\ee
and separating the
variables we get the radial and angular equations
\be
\left\{ \begin{array}{l}
\dfrac{1}{r}\dfrac{\partial}{\partial r}\left(r \dfrac{\partial
R}{\partial r} \right) - \dfrac{\mu}{r^2}R + \dfrac{2}{s^2}({\cal
E}-V(r))R = 0
\\
\\
\P^2\Phi = -\mu \Phi
\end{array}
\right.
\label{radang}
\ee
where we suppose $\mu \in \mathbb{R}$.
\section{The radial equation}
\setcounter{equation}{0}
The radial equation is quite similar to the standard one of the
hydrogen atom theory, apart of course for the radial part of the
Laplacian, which is here 2-dimensional. For its solution we shall
therefore use standard techniques (see e.g. \cite{landau}).\\
Let's now look for the asymptotic behaviour of $R(r)$. We ask
$R(r)$ to be finite everywhere including $r=0$. Under the
hypothesis
\be
\lim_{r \rightarrow 0} V(r) r^2 = 0\, ,
\ee
which is fulfilled by $V(r)=-GM/r$, the radial equation for $r
\rightarrow 0$ becomes
\be
r\frac{\partial}{\partial r}\left(r
\frac{\partial R}{\partial r} \right) - \mu R = 0  \, . \label{rr}
\ee
We seek $R(r)$ in the form of a power series and we retain only
the first term for small $r$. That is, we put $R(r)=kr^t$ for $r
\rightarrow 0$. Substituting this in the equation (\ref{rr}) we
find
\be
t^2 =\mu\,.
\ee
We want $R(r)$ real, therefore $t$ must
be real, and $\mu \geq 0$. So we have two roots
\be
t_1&=&-\sqrt{\mu} \nonumber \\
t_2&=&+\sqrt{\mu}\,.
\ee
But $t_1\leq 0$, hence $r^{t_1}
\rightarrow \infty$ for $r \rightarrow 0$. So $t_1$ does not
yields a $R(r)$ finite near the origin, and must be discarded. The
only acceptable solution is $t_2=+\sqrt{\mu} \geq 0$. Therefore we put
$R(r)\sim k r^{t_2}$ for $r \rightarrow 0$.
For the
Newtonian potential $V(r)=-GM/r$ the radial equation reads
\be
 \frac{\partial^2 R}{\partial r^2} + \frac{1}{r} \frac{\partial R}{\partial r}
 - \frac{\mu}{r^2}R +
\frac{2}{s^2}\left({\cal E}+\frac{GM}{r}\right)R = 0\,. \nonumber
\ee
We choose as natural units for mass, length, and energy (per unit mass),
respectively,
\be
M, \quad \quad \frac{s^2}{GM}, \quad \quad
\frac{G^2M^2}{s^2}
\ee
so that the radial equation can be
rewritten as
\be \frac{\partial^2 R}{\partial r^2} + \frac{1}{r}
\frac{\partial R}{\partial r} - \frac{\mu}{r^2}R + 2\left({\cal
E}+\frac{1}{r}\right)R = 0\,.
\label{rnew}
\ee
To study the
discrete spectrum (bound orbits, ${\cal E} < 0$), we introduce, in
place of $\cal E$ and $r$, the variables
\be
n=\frac{1}{\sqrt{-2\,{\cal E}}} \quad \quad {\rm and} \quad \quad
\rho=\frac{2}{n}\,r = 2\sqrt{-2\,{\cal E}}\,r
\ee
with ${\cal E}<0$, $n>0$, and their inverse relations
\be
{\cal E}=-\frac{1}{2n^2}, \quad \quad r=\frac{n}{2}\,\rho\,.
\ee
Equation (\ref{rnew}) then becomes
\be
\frac{\partial^2
R}{\partial \rho^2} + \frac{1}{\rho} \frac{\partial R}{\partial
\rho} + \left(-\frac{1}{4}+\frac{n}{\rho} -
\frac{\mu}{\rho^2}\right)R = 0\,.
\label{rrho}
\ee
We already know
that $R(\rho)\sim\rho^{t_2}$ for $\rho\rightarrow 0$. If now we
take $\rho\rightarrow +\infty$, then eq. (\ref{rrho})
reads\footnote{Since the normalization condition (\ref{norm}) implies
$R(\rho)\rightarrow 0$, then $R'(\rho)\rightarrow 0$, for
$\rho \rightarrow +\infty$.}
\be
\frac{\partial^2 R}{\partial \rho^2} -\frac{1}{4}R=0
\ee
whose solutions are $e^{\pm\rho/2}$. We want $R(\rho)\rightarrow 0$ for
$\rho \rightarrow +\infty$, therefore we must choose the second,
$R(\rho)\sim e^{-\rho/2}$ for $\rho \rightarrow +\infty$.\\
If now we make the substitution
\be R(\rho) = \rho^{t_2}\,
e^{-\rho/2}\,w(\rho)
\ee
then the eq. (\ref{rrho}) becomes
\be
\rho\, w'' + (2\,t_2 +1 -\rho)\,w' + (n-t_2-\frac{1}{2})\,w =0.\nonumber
\ee
We look for a solution of this equation which diverges at
infinity no more rapidly than a finite power of $\rho$, while for
$\rho \rightarrow 0$ we should have $w \rightarrow w_o$ finite.
Such a solution is the {\em confluent hypergeometric function}
(see Appendix)
\be w(\rho)= F(\alpha, \gamma, \rho)
= F(t_2 +\frac{1}{2} - n,\,\, 2\,t_2+1,\,\, \rho).
\label{wF}
\ee
In
particular, it behaves as a polynomial ($w\rightarrow \rho^p$ for
$\rho \rightarrow +\infty$) only if $\alpha =-N$ with $N\geq0$
integer. Thus
\be t_2+\frac{1}{2} -n = -N \quad \quad
\Longrightarrow \quad \quad n=t_2+\frac{1}{2} + N\,,
\ee
where $N=0,1,2,3,\dots$.
\section{The angular equation and the spectrum of ${\cal E}$}
\setcounter{equation}{0}
We need now to know what is $t_2=\sqrt{\mu}$. To this aim, we solve the angular equation
\be
\P^2\Phi=-\mu\Phi \quad \quad \quad \mu\geq0\,.
\label{P2}
\ee
We take $\Phi\in{\cal L}^2([0, 2\pi])$ and we know that an
orthonormal basis in ${\cal L}^2([0, 2\pi])$ is
$e^{im\phi}/\sqrt{2\pi}, m\in \mathbb{Z}$.
From the definition
(\ref{defP}) we immediately get
\be
\P^2 e^{im\phi}=-e^{2\lambda m}e^{im\phi}
\ee
which means that
\begin{itemize}
\item the eigenvectors of $\P^2$ are $u_m=e^{im\phi}/\sqrt{2\pi}$;
\item the eigenvalues of $\P^2$ are $\{-e^{2\lambda m}\}_{m\in\mathbb{Z}}$, and
therefore $\mu_m=e^{2\lambda m}$, $m=0, \pm1, \pm2, ...$.
\end{itemize}
Hence we have
\be
t_2 = \sqrt{\mu}= e^{\lambda m} \quad \quad  m=0, \pm1, \pm2, ...
\ee
and then
\be
n=t_2+\frac{1}{2} +N = e^{\lambda m}+\frac{1}{2} +N
\ee
with $N=0,1,2,3,...$, $m=0, \pm1, \pm2, ...$.\\
Finally, we are able to write down the spectrum of the energy (per unit mass)
which in ordinary units reads
\be
{\cal E} = -\frac{1}{2}\left(\frac{GM}{s}\right)^2\frac{1}{(e^{\lambda m}+1/2+N)^2}
 = -\frac{1}{2}\,\frac{GM}{a}\,\frac{1}{(e^{\lambda m}+1/2+N)^2}
\label{eltb}
\ee
with $N=0,1,2,3,...$, $m=0, \pm1, \pm2, ...$\,.\\
A couple of considerations are now in order:
\begin{itemize}
\item For $N$ fixed, and large positive $m$, the levels behave like
\be
{\cal E}\sim-\frac{1}{2}\left(\frac{GM}{s}\right)^2\frac{1}{e^{2\lambda m}}\, ,
\ee
i.e. we recover the formula (\ref{etbam}) obtained from the model {\em a la} Bohr.
\item Clearly, the best fit between the relations
(\ref{eltb}) and (\ref{etbam}) (the relation (\ref{etbam}) is directly given,
in practice, by the observation) is obtained for $N=0$.
The sequence $N=0$ and $m=1,2,3,...$
is called the {\em principal sequence}. The formula (\ref{eltb}) seems
to suggest the existence of other sequences. In particular, we shall see that the sequence
\be
N=0 \quad \quad {\rm and} \quad \quad m=0,-1,-2,-3,...\,.
\ee
can be interpreted a system of rings (see section \ref{secrings}).
\end{itemize}
\section{Mean value of $r$}
\setcounter{equation}{0}
In order to complete our analysis of the solution of eq.
(\ref{radang}), we want to compute the mean values taken by the
variable $r$ in various eigenstates. We are particularly
interested in the eigenstates of the principal sequence, $N=0,
\quad m=1,2,3,..$, which is the one matching the observational
data in the closest way. For $\psi=R(r)\Phi(\phi)$, the
normalization condition (\ref{norm}) holds, and the $\Phi
(\phi)=e^{im\phi}/\sqrt{2\pi}$ are already normalized to unity.
Therefore we are left with the condition on $R(r)$ only
\be
1=\int_0^\infty dr \,r \,R(r)^2\,.
\ee
First, we find the correct normalization constant
for $R(\rho)$ by translating the normalization condition for
$R(r)$ into the one for $R(\rho)$. Writing
\be
R(\rho) = A\,\rho^{t_2}\, e^{-\rho/2}\,w(\rho)\,,
\ee
we have (reminding $\rho=2r/n$)
\be
1 = \int_0^\infty dr\, r\, R(r)^2 = \frac{n^2}{4} A^2 \int_0^\infty d\rho\,
\rho^{2t_2 + 1}\,e^{-\rho}\,[w(\rho)]^2\,.
\ee
Since we are mainly
interested in the mean values of $r$ for the eigenstates belonging
to the principal sequence, ($N=0, \quad m=1,2,3,..$), we set
$N=n-t_2-1/2=0$. This means $n=t_2+1/2$ and
\be
w(\rho)=F(0, \,2\,t_2 + 1, \,\rho) =1
\ee
This simplify very much the calculation of
the integral. Finally we have
\be
A^2 = \frac{2}{n^3\,\Gamma (2\,t_2+1)}
\label{A}
\ee
where $\Gamma$ is the Euler' $\Gamma$-function, and we used $2\,n=2\,t_2+1$ and $\Gamma (x+1)=x\Gamma(x)$.\\
We can now compute the mean value of $r$:
\be
\bar{r} = \int_0^\infty dr\, r^2\, R(r)^2
 = \frac{n^3}{8} A^2 \int_0^\infty d\rho\, \rho^{2t_2 + 2}\,e^{-\rho}\,[w(\rho)]^2\,.
\ee
Again, we are interested in the principal sequence. Setting $N=0$ and $w(\rho)=1$, and using
for $A$ the value just obtained in (\ref{A}), we have
\be
\bar{r} = \frac{n^3}{8} A^2 \int_0^\infty d\rho\, \rho^{2t_2 + 2}\,e^{-\rho} =
\frac{n^3}{8} A^2\, \Gamma(2\,t_2+3)=
n^2+\frac{n}{2}\,.
\ee
Since $N=0$, then $n=t_2+1/2=e^{\lambda m}+1/2$. Therefore
\be
\bar{r}=n^2+\frac{n}{2} = (e^{\lambda
m}+1/2)^2+\frac{1}{2}(e^{\lambda m}+1/2)
\label{rings}
\ee
Restoring the ordinary units, we have as the mean value of $r$, for large and positive $m$,
\be
 \bar{r} \sim \frac{s^2}{GM}e^{2\lambda m} = a\, e^{2\lambda m}
\ee
which agrees with the Bohr model developed in section \ref{Sec4}.\\
We see here that the agreement with the observed T-B rule holds only for large $m$.
The error is particularly relevant for the internal planets, a feature unluckly shared with all
the "quantum" models of the T-B rule. In some of them (e.g., those of Reinisch, Nottale, Oliveira)
non consecutive integers have been chosen in order to recover contact with the observation.
Others models (Albeverio) also cite
a necessary breakdown of the "quantum" T-B rule near the sun, because effects different
from the gravitational ones become important. We tend to support this last explanation.
\section{Description of the rings}
\setcounter{equation}{0}
\label{secrings}
It is tempting to speculate on other possible sequences, in
particular the one $N=0$ and $m=0,-1,-2,-3,...$\,.
First, we note that for $N=0$ the mean value $\bar{r}$ is still given by the relation
(\ref{rings}), or in ordinary units
\be
\bar{r} = a\left[(e^{\lambda m}+1/2)^2 + \frac{1}{2}(e^{\lambda m}+1/2)\right]
\label{rings2}
\ee
where $a=s^2/(GM)$. A we said, the formula (\ref{rings2}) matches the phenomenological
formula (\ref{tbl}) only for large, positive $m$. And this could be expected as due to
the "quantum" origin (i.e. from a wave equation) of the formula (\ref{rings2}).
But the interesting feature of the equation (\ref{rings2}) is that it can be now considered
also for $m=0,-1,-2,-3,...$. In the limit $m\rightarrow -\infty$ we have for (\ref{rings2})
\be
\bar{r}\longrightarrow \frac{1}{2}\,\, a \, ,
\label{radring}
\ee
contrary to the limit given by the phenomenological rule (\ref{tbl}), which predicts
$\bar{r}\rightarrow 0$ for $n\rightarrow -\infty$.
The sequence $m=0,-1,-2,-3,...$ would in such a way correspond to a system of
permitted concentric orbits, accumulating on the limit orbit $\bar{r}=a/2$: clearly,
a system of rings.
We can check the predictive ability of the relation (\ref{radring}) using the planets
equipped with a system of rings: Jupiter, Saturn, Uranus.
The procedure is the following.

First, we calculate the parameter $a$ from the phenomenological Titius-Bode rule, using the
radius $r_1$ of the first satellite orbit:
\be
a=\frac{r_1}{e^{2\lambda}}.
\ee

Then we calculate the radius that the inner ring should have:
\be
R_{in-ring}=\frac{1}{2}\,a=\frac{r_1}{2e^{2\lambda}}.
\ee
Finally, we compare the above relation with the observational data, using the reverse form
\be
2\,e^{2\lambda}\,R_{in-ring}=r_1.
\ee
With the data given, for example, in \cite{Sat}
we can write, for:\\

{\bf Jupiter:}
\be
2\,e^{2\lambda}\,R_{in-ring} = 2 \cdot 1.7277 \cdot 122\,500 = 423\,280 \quad \textrm{km} \nonumber
\ee
This value agrees almost perfectly (within an error of less than $1\,\%$) with the radius
of the orbit of Io, which is $421\,600 \,\, \textrm{km}$.
As first ring, we considered the ring "Main". Besides, we considered Io as the first satellite
(as regard the distance from Jupiter), instead of Metis, Adrastea, Amalthea or Thebe.
This because the latter satellites have sizes of, at most, $100 \,\, \textrm{km}$ and
masses which
are $10^{-4}$ - $10^{-6}$ the mass of Io.
This choice takes into account the well known fact that the Titius-Bode rule works well for
quite large and quite massive objects. For example, it does not work for comets or light
asteroids. This criterium will be adopted also in the forthcoming considerations about
the Saturn and Uranus systems.\\

{\bf Saturn:}
\be
2\,e^{2\lambda}\,R_{in-ring} = 2 \cdot 1.5967 \cdot 66\,000 = 210\,700 \quad \textrm{km} \nonumber
%\label{*}
\ee
Here we have used the radius of the ring "D", the inner one. We see that the number
obtained lies half a way in between the orbits of Mimas ($185\,520\,\,\textrm{km}$)
and Enceladus
($238\,020\,\,\textrm{km}$), which are the first two "big" satellites. The error is in
the range
of $11.5\,\%\,-\,13.5\,\%$. If we use the radius of the ring "C", namely
$R_{in-ring}=74\,500\,\,\textrm{km}$, we get for $r_1=237\,900\,\,\textrm{km}$,
which agrees
almost perfectly with the orbital radius of Enceladus (less than $1\,\%$ error).
However, Enceladus is not the first satellite (as regard the distance from Saturn)
but only the second. Here also, as
already done for Jupiter, we have discarded the too light bodies, and the satellites
discovered only with spacecrafts (therefore, very small). The reasons are the same
as in the above.\\

{\bf Uranus:}
\be
2\,e^{2\lambda}\,R_{in-ring} = 2 \cdot 1.4662 \cdot 41\,840 = 122\,690 \quad \textrm{km} \nonumber
%\label{*}
\ee
This value agrees with the orbital radius of Miranda ($129\,780\,\,\textrm{km}$),
the first "big" moon considered, within an error of $5.5\,\%$. Here we considered
as first inner ring the ring "6". Miranda is the first satellite with relevant mass
and size. In fact it was discovered from the Earth by Kuiper in 1948. If we use the
ring "Alpha" as inner ring, we get for the radius of the first satellite orbit
$r_1=131\,140\,\,\textrm{km}$, in agreement within $1\,\%$ with the Miranda orbital radius.\\

Even if the agreement between the predicted inner radius of the rings and the
observational data is not perfect (however with errors around $10\,\%$ or less), and the
statistics of only three cases is really poor, nevertheless this "prediction" seems
to corroborate the quantum-like model presented in this paper. On the other hand,
it should be noted that these errors are of the same order of magnitude of those affecting the
phenomenological Titius-Bode rule, therefore acceptable in this context.
\section{Permitted orbits, dissipation and gravity}
\setcounter{equation}{0}
One of the main objections that it is possible to rise against the
existence of permitted discrete orbits in a planetary system is
the following: It is a common experience, in this era of space
travels, that a satellite can be put in any orbit we wish around
the Earth, the Sun, or any other planet. Why therefore there
should exist stable {\em permitted} orbits? In what sense they are
"permitted"? How they are reached? Answers to these questions are suggested
by the concepts of (chaotic) {\em dissipation} and
{\em limit cycles}, emphasized in papers
\cite{hooft1} (see also further references therein). Although in
the proto-planetary nebula dust, particles and other bodies can
be found at any distance from the central body, after a huge
amount of time, friction, viscosity, and chaotic gravitational effects (for example planetary
resonances) produce
a dissipation of the initial state and bring matter to stabilize
in several orbits, the limit cycles, where particles and dust
aggregate to form planets. It is in this way that we can speak
about "permitted orbits": they are the "limit cycles" of the
dissipative processes started in the primitive nebula.\\
Of course, it is in principle possible to put, today, a body in
any orbit we wish. But if we wait for a time of the order of
$5\cdot 10^9$ years, and if such body has a sufficient mass, it is
likely that we finally find it in one of the permitted limit
cycles.\\
The present model is therefore somehow complementary to the Classical Mechanics explanation
that orbits in between the observed one are unstable, due to three-body forces
(gravitational resonances). The quantum-like language of a wave equation is a tool to
describe these collective effects of classical forces acting on unstable orbits.
This "quantum" collective behaviour is an emerging phenomenon from an underlying classical
deterministic (even if chaotic) dynamics. The description through a wave function is moreover
very economical: it enables to get, at once, the satellite orbits and the rings.
\section{Conclusions}
\setcounter{equation}{0}
In this paper we have shown that a wave equation is capable, under certain restrictive
conditions, to describe some basic properties of the planetary systems, namely the
law of the distances of the planets (or satellites) from the central body. The wave
equation adopted for this scope is a deformation of the
Schr\"{o}dinger equation. \\
We have been forced to choose a deformed Schr\"{o}dinger equation by the
analogy between the mechanism devised in \cite{hooft1} to produce quantization at atomic
level via dissipation, and the chaotic dissipation occurred during the history of the
proto-planetary nebulae. An dissipation mechanism could have been
at work (of course, on much larger time scales) during the evolution of the planetary
systems.
From the primitive nebula, where all the orbits were filled by dust and rubble, we
arrive, after $5$ billion years of evolution, to the "quantized" orbits of today.\\
Of course, having marked the analogies, also the evident differences must be underlined.
We do not have quantum jumps in the Solar System, or quantum interferences
between planets, neither quantum superpositions nor zero-point energy. A planetary
system {\em is not}, of course, a quantum system.
On the contrary, we have shown that a deformation of the Schr\"{o}dinger equation
(one of the basic tools of Quantum Mechanics) seems to be able to play a role also
in the description of some quantum-like features of the planetary systems.
Although indirectly, this quantum-like description seems also to corroborate
the 't Hooft ideas on the origin of quantization from dissipation.
\section*{Acknowledgement} \nonumber
The author gratefully thanks for enlightening discussions M.
Blasone, E. Gozzi, P. Jizba, H. Kleinert, G. Vitiello. Thanks also
to G. Immirzi for having drawn the author's attention on the first
paper in Ref. \cite{hooft1}, and to M. Arpino for providing some literature.
This work has been written at CENTRA - Instituto Superior Tecnico - Lisbon.
The author is presently supported by the JSPS Research Fellowship P06782.
\section*{Appendix}
\setcounter{equation}{0}
\subsection*{Linearity, self-adjointness and explicit form of $\p_\phi=-is\P$}
$\P$ is defined on the orthonormal basis
$u_m=e^{im\phi}/\sqrt{2\pi}$ of ${\cal L}^2([0, 2\pi])$ as
\be
\P\,e^{im\phi}:=i\,e^{m\lambda}\,e^{im\phi}\,.\nonumber
\ee
Defining
$\p_\phi=-is\P$ we have
\be
\p_\phi\,e^{im\phi}=se^{m\lambda}e^{im\phi}\,.\nonumber
\ee
Hence
$u_m=e^{im\phi}/\sqrt{2\pi}$ are the eigenvectors of $\p_\phi$
with the eigenvalues $\mu_m = se^{m\lambda}$.\\
Moreover we define $\P$ to be linear by stipulating
\be
\left\{\begin{array}{l}
\P(e^{in\phi}+e^{im\phi}):= i\,e^{n\lambda}\,e^{in\phi}+i\,e^{m\lambda}\,e^{im\phi}
=\P(e^{in\phi})+\P(e^{im\phi}) \nonumber
\\
\\
\P(\alpha
e^{in\phi}):=i\alpha\,e^{n\lambda}\,e^{in\phi}=\alpha\P(e^{in\phi})\,.
\end{array}
\right.
\ee
Being $\P$ linear on an orthonormal basis, then $\P$
is linear all over ${\cal L}^2([0, 2\pi])$.\\
We remind also the \\
Theorem: {\em If a linear operator is self-adjoint on an
orthonormal basis of a Hilbert space ${\cal H}$,
then it is self-adjoint over all ${\cal H}$}.\\
Therefore we have simply to show that $\p_\phi$ is self-adjoint on
the basis $u_m$. In fact we have
\be
\langle\p_\phi u_m|u_n\rangle =
\int_0^{2\pi}d\phi\,(\p_\phi u_m)^*u_n
=\frac{se^{m\lambda}}{2\pi}\int_0^{2\pi}d\phi\,
e^{-im\phi}e^{in\phi} = se^{m\lambda}\delta_{mn} \nonumber
\ee
and
\be
\langle u_m|\p_\phi u_n\rangle = \int_0^{2\pi}d\phi\,u_m^*(\p_\phi u_n)
=
\frac{se^{n\lambda}}{2\pi}\int_0^{2\pi}d\phi\,
e^{-im\phi}e^{in\phi} = se^{n\lambda}\delta_{mn}\,. \nonumber
\ee
\\
\\
Besides, we check the identity
\be
\P
=ie^{-i\lambda\partial_\phi}\,.\nonumber
\ee
In fact, since
\be
-i\lambda\partial_\phi e^{im\phi} = \lambda m e^{im\phi} \,,\nonumber
\ee
we have
\be
e^{-i\lambda\partial_\phi}\,e^{im\phi}
&=&\left(1+(-i\lambda)\partial_\phi+...+
\frac{1}{n!}(-i\lambda)^n\partial_\phi^n +...\right)\cdot
e^{im\phi}\nonumber \\
&=& \left(1+\lambda m +...+
\frac{\lambda^n m^n}{n!}+...\right)\cdot
e^{im\phi}=e^{m\lambda}e^{im\phi}\,.\nonumber
\ee
Hence, the thesis.
\subsection*{Confluent hypergeometric function}
The {\em confluent hypergeometric function} is defined via the
series
\be
F(\alpha, \gamma, z)= 1
+\frac{\alpha}{\gamma}\frac{z}{1!}+
\frac{\alpha(\alpha+1)}{\gamma(\gamma+1)}\frac{z^2}{2!}+...\,,\label{d1}
\ee
which converges for all finite $z$; the parameter $\alpha$ is any
number in $\mathbb{C}$, and the parameter $\gamma$ must be
different from zero and from any negative integer. If $\alpha$ is a
negative integer (or zero), the function $F(\alpha, \gamma, z)$
becomes a polynomial of degree $|\alpha|$. \\
The function $F(\alpha, \gamma, z)$ is a solution of the
differential equation
\be z u'' + (\gamma - z)u' - \alpha u = 0\,,
\label{d2}
\ee
as can be directly checked.\\
With the substitution $u=z^{1-\gamma}u_1$, eq.(\ref{d2})
transforms into
\be
z u_1'' + (2-\gamma - z)u_1' -(\alpha-\gamma+1) u_1 = 0\,.\nonumber
\ee
From here we see that, for a non integer $\gamma$,
eq.(\ref{d2}) admits also the integral
\be
z^{1-\gamma}F(\alpha-\gamma+1,\,\, 2-\gamma,\,\, z) \nonumber
\ee
which is linearly
independent from (\ref{d1}), so that the general solution of
eq.(\ref{d2}) has the form
\be
u=c_1 F(\alpha, \gamma, z)\, + \,c_2\,
z^{1-\gamma}F(\alpha-\gamma+1, \,\,2-\gamma,\,\, z)\,.\nonumber
\ee
The second
term, contrary to the first, has a singularity in $z=0$ when $\gamma>1$.
\end{document}